\documentclass[prd,preprintnumbers,longbibliography,
superscriptaddress,tightenlines,twocolumn,nofootinbib
]{revtex4-1}

\usepackage[utf8]{inputenc}
\usepackage{slashed}
\usepackage{amsmath}
\usepackage{amsfonts}
\usepackage{dsfont}
\usepackage{xcolor}
\usepackage{tikz}
\usepackage{hyperref}
\usepackage{color}
\usepackage{orcidlink}
\usepackage{xspace}
\usepackage{placeins}

\usepackage{ulem}

\usetikzlibrary{decorations.shapes}
\usetikzlibrary{decorations.shapes,shapes.geometric} 

\graphicspath{{./Figures/}} 

\newcommand{\chem}{\mu}
\newcommand{\dd}{\mathrm{d}}

\def \lr#1{\left( #1 \right)}

  \newcommand{\App}[1]{App.\,\ref{app:#1}}

  \newcommand{\Eq}[1]{Eq.\,\eqref{eq:#1}}

  \newcommand{\Fig}[1]{Fig.\,\ref{fig:#1}}
  
  \newcommand{\FIG}[1]{Figure\,\ref{fig:#1}}

  \def\Zthree{\ensuremath{\mathbb{Z}_3}\xspace}
  \newcommand{\U}[1]{\ensuremath{{\rm U}(#1)}\xspace}

  \newcommand{\SUf}[1]{\ensuremath{{\rm SU}(#1)_f}\xspace}
  \newcommand{\SU}[1]{\ensuremath{{\rm SU}(#1)}\xspace}
  \newcommand{\SO}[1]{\ensuremath{{\rm SO}(#1)}\xspace}

  \def\bchi{\boldsymbol{\chi}}

\tikzset{decorate sep/.style 2 args=
{decorate,decoration={shape backgrounds,shape=circle,shape size=#1,shape sep=#2}}}
\tikzset{decorate sepbis/.style 2 args=
{decorate,decoration={shape backgrounds,shape=rectangle,shape size=#1,shape sep=#2}}}
\tikzset{decorate septris/.style 2 args=
{decorate,decoration={shape backgrounds,shape=diamond,shape size=#1,shape sep=#2}}}

\begin{document}

\title{Mass gaps of a \Zthree gauge theory with three fermion flavors in $1+1$ dimensions}

\author{Adrien Florio \,\orcidlink{0000-0002-7276-4515}}\email{aflorio@bnl.gov}
\affiliation{Department of Physics, Brookhaven National Laboratory, Upton, NY 11973}

\author{Andreas Weichselbaum \,\orcidlink{0000-0002-5832-3908}}
\email{weichselbaum@bnl.gov}
\affiliation{Department of Condensed Matter Physics and Materials
Science, Brookhaven National Laboratory, Upton, NY 11973-5000, USA}

\author{Semeon Valgushev \,\orcidlink{0000-0002-4306-1423}}
\email{semeonv@iastate.edu}
\affiliation{Department of Physics and Astronomy, Iowa State University, Ames, IA, 50011, USA}
\author{Robert D. Pisarski \,\orcidlink{0000-0002-7862-4759}}\email{pisarski@bnl.gov}
\affiliation{Department of Physics, Brookhaven National Laboratory, Upton, NY 11973}

\date{}

\begin{abstract}

We consider a \Zthree gauge theory coupled to
three degenerate massive flavors of fermions, which we term ``QZD".
The spectrum can be computed in $1+1$ dimensions using
tensor networks.  In weak coupling the spectrum is
that of the expected mesons and baryons, although the corrections in
weak coupling are nontrivial, analogous to those of non-relativistic
QED in $1+1$ dimensions.  In strong coupling, besides the usual
baryon, the singlet meson is a baryon anti-baryon state.  For
two special values of the coupling constant, the lightest baryon
is degenerate with the lightest octet meson, and the lightest
singlet meson, respectively.

\end{abstract}

\date{\today}

\maketitle

\section{Introduction}
\label{sec:introduction}

How confinement in gauge theories produces a non-trivial mass spectrum is a problem of fundamental importance for a wide variety of problems, from Quantum ChromoDynamics (QCD) in the strong interactions \cite{Wilson:1974sk,Kogut:1974sn}, to numerous systems in condensed matter \cite{Fradkin:2013sab}.

The simplest examples of confining gauge theories are of course
in the fewest number of spacetime dimensions, which is $1+1$.
The iconic examples are fermions coupled to an Abelian gauge
theory, which is the Schwinger model
\cite{Schwinger:1962tp,Banks:1975gq,Coleman:1975pw,Coleman:1976uz,Manton:1985jm}
and $N_f$ flavors of quarks coupled to a \SU{N_c} non-Abelian
gauge theory, with $N_c$ the number of colors.
For $N_f \ll N_c \rightarrow \infty$, this is the
't~Hooft model
\cite{tHooft:1974pnl,Callan:1975ps,Fonseca:2006au,Fateev:2009jf,Ziyatdinov:2010vg,Zubov:2015ura}.

As an Abelian theory the Schwinger model is especially useful.  For a single, massless fermion, 
Schwinger showed that the only gauge invariant state is a single free, massive boson \cite{Schwinger:1962tp}.  When the fermions are massive, however, there is 
an infinite number of gauge invariant pairs of fermions and anti-fermions.  These obviously do not carry fermion number, and are a type of meson. When their mass is large, Coleman computed the number of mesons semi-classically \cite{Coleman:1976uz}. 

While classical computers can be used to numerically compute many properties of field theories, there are some aspects --- notably the evolution in real time, or theories with a sign problem --- for which quantum computers are necessary.  This requires controlling the Hilbert space of a field theory, which even with a lattice regularization is exponentially large.  In $1+1$ dimensions though, polynomial approximations have been developed, as matrix product states (MPS) efficiently represent the ground states of gapped systems \cite{Buyens:2013yza,Banuls:2013jaa,Cirac:2020obd}. 
Studies of the Schwinger model on quantum computers include Refs. \cite{Hauke:2013jga,Rico:2013qya,Pichler:2015yqa,Zohar:2015hwa,Martinez:2016yna,Muschik:2016tws,Klco:2018kyo,Banuls:2019bmf,Surace:2019dtp,Banuls:2019rao,Yang:2020yer}.
Other properties analyzed include how mesons scatter
\cite{Rigobello:2021fxw,Rigobello:2023ype}, thermalization
\cite{Desaules:2022ibp,Chanda:2019fiu}, string breaking
\cite{Pichler:2015yqa,Kuhn:2015zqa,Magnifico:2019kyj,deJong:2021wsd,Lee:2023urk},
entanglement production in jets \cite{Florio:2023dke} and
the dynamics in $\theta$-vacuum \cite{Zache:2018cqq, Kharzeev:2020kgc,Ikeda:2020agk}. 

In the massive Schwinger model the only states which survive confinement are mesons.
It would be useful to study models where confinement produces states which do carry net fermion number, analogous to baryons in QCD.  

There are several such models in $1+1$ dimensions.  As a \SU{N_c} gauge theory, the 't Hooft model has baryons, but their properties are
opaque \cite{Witten:1979kh,Steinhardt:1980ry,Bringoltz:2009ym}.  A \SU{N_c} gauge theory coupled to $N_f$ light flavors of quarks can be analyzed using conformal field theory, as a type of Wess-Zumino-Novikov-Witten model \cite{Lajer:2021kcz}; it behaves in a manner characteristic of such two dimensional theories.
Rico {\it et al.} \cite{Rico:2018pas} studied a \SO{3} model in which both the quarks and the gluons lie in the adjoint representation, and so a quark and a gluon can directly combine  to form a gauge invariant fermion.  This is like a \SU{N_c} gauge theory coupled to quarks in the adjoint representation, instead of the fundamental representation as in QCD. Lastly, Farrell {\it et al.}
\cite{Farrell:2022wyt} directly integrated out the \SU{N_c} gauge fields, which is possible in $1+1$ dimensions, 
to obtain the mass spectrum for \SU{3} gauge fields coupled to two massive flavors. 

While these models are all useful, we wish to study a simpler model where fermions emerge as gauge invariant states.
Before doing so, it is necessary to explain in detail why 
in the Schwinger model, a single, massive flavor
has no gauge invariant states with net fermion number.  In Minkowski spacetime, the total Hamiltonian is
\begin{equation}
    H = \int\dd x \left[
        \overline{\psi}\left(-i\gamma^1\partial_1
      + \gamma^1A_1+m\right)\psi \right
    ] + \tfrac{g^2}{2} E^2  \ ,
\label{eq:H:continuum}
\end{equation}
where $E$ is the electric field operator, and $A_1$ the conjugate gauge potential.  Gauge invariance requires that we 
impose Gauss's law,  
\begin{equation}
    \partial_1 E = \overline\psi \gamma^0 \psi \ .
\end{equation}
The right hand side is just the charge density for
the fermion field, which for a single flavor, is {\it identical}
to the density for fermion number.
(To represent Wilson loops it is necessary to add an external charge density to the right hand side, which
manifestly extends the Hilbert space
\cite{Gervais:1978kn,Pisarski:2022cuo,Kaplan:2023fbl}.)
Computing the total electric charge, 
$Q_{\rm tot}$, Gauss's law gives
\begin{equation}
    Q_{\rm tot}
    =\int\dd x\, \partial_1 E = E(\infty) -E(-\infty) \ . 
\end{equation}
For the system to be well defined in the limit of infinite
volume, we require that there is no net electric field,
$E(\infty) =E(-\infty)$, and the total electric charge vanishes,
$Q_{\rm tot}=0$ 
Further, since for a single flavor the total fermion number
equals the total charge $N_{\rm tot} =0$
as well, where by $N$ we mean the number of 
particles relative to half-filling, or equivalently,
relative to the ground state.

Thus in a \U{1} theory in $1+1$ dimensions, 
for a single flavor Gauss's law prevents us from introducing
{\it any} net fermion number.  In short, 
the global \U{1} symmetry of fermion number 
is already part of the \U{1} gauge symmetry.  

This can be seen explicitly by trying to introduce a chemical potential for fermion number, $\mu$.
In the Hamiltonian formalism all thermodynamics quantities follow from the partition function, 
\begin{equation}
    Z(T,\mu)=\mathrm{Tr} \left( 
    e^{- (H - \chem N_{\rm tot})/T}
    \right)
\; ,
\label{eq:partition_function}
\end{equation}
where the trace is over all physical states.
Physical states, though, must obey Gauss's law.
For \U{1}, this enforces $Q_{\rm tot} = N_{\rm tot}=0$,
and consequently, that
the partition function is independent of $\mu$, $Z(T,\mu)=Z(T,0)$.

This is can also be seen directly using the Lagrangian formalism.
For a single flavor, $\mu \neq 0$ can 
be eliminated simply by shifting the time like component of
the vector potential by an imaginary constant, 
$A_0\rightarrow A_0-i\chem/g$ \cite{Dumitru:2005ng}. 

With two or more flavors, then clearly one can introduce a fermion number for one flavor relative to those for the others.
This is evident for two flavors, which we call up, $u$, and down, $d$.  Then a net electric charge from an excess of $u$ fermions over $u$ anti-fermions can be precisely cancelled by an excess of down anti-fermions, $\overline{d}$, over $d$ fermions.
This is obviously just a chemical potential for isospin between the up and down quarks.  While an isospin chemical potential exhibits interesting phenomena, such as spatially varying phases \cite{Narayanan:2012qf,Lohmayer:2013eka,Banuls:2016gid},
it still leaves us bereft of gauge invariant fermions.

A simple model where there are both gauge invariant fermions and bosons was proposed in Ref. \cite{Pisarski:2021aoz}.  Consider a 
\Zthree gauge theory coupled to
three, degenerate massive flavors of fermions, adding strange, $s$, to $u$ and $d$.  
By the Fermi exclusion principle, we cannot put two identical fermions at the same point in space, since $u^2=0$, {\it etc.}
This is unlike QCD, where three quarks of the same flavor can sit on the same point in space, as long as they each carry a different color; for example, in QCD the $\Omega$ baryon is $sss$.  Assuming that the \Zthree gauge theory confines, the only way to put fermions at the same point in space is if they have different flavors.  
Thus the simplest singlet under the \Zthree gauge group is $uds$,
which is like the $\Lambda$ baryon in QCD.

Confinement also produces mesons in this $\mathbb{Z}_3^{\; 3}$ theory, but these are simple to understand.  Since there are three
degenerate flavors, we can form $\mathbb{Z}_3$ singlets in two ways.  There is a flavor singlet, 
\begin{equation}
    \eta' = \sum_{f = 1}^3 \overline{\psi}^f \psi^f\ ,
\end{equation}
and a flavor octet,
\begin{equation}
\pi^A = \sum_{f,g=1}^3 \overline{\psi}^f t^A_{f g}\,  \psi^g \ ;
\end{equation}
$f$ and $g$ are indices for the fundamental representation of
flavor, $f,g = 1,2,3$, while $t^A_{f g}$ is a \SU{3} flavor matrix in the adjoint representation, $A = 1\ldots 8$.  As suggested by the notation,
the singlet meson is like the $\eta'$ meson in QCD, while the octet multiplet $\pi^A$ is analogous to the $\pi$, $K$, and $\eta$ mesons.

Thus we have a model which has gauge invariant singlets
which are both fermions (baryons) and bosons (mesons).
To avoid the 
the subtleties and complications of chiral symmetry breaking in
two spacetime dimensions, we take the fermions to all have the same, nonzero mass. 

In this paper we study the mass spectrum of the lightest states
of this $\mathbb{Z}_3^{\; 3}$ theory as a function
of the coupling constant on the lattice.  First we discuss the theory on a lattice, and how to obtain a \Zthree gauge theory from the spontaneous breaking of a \U{1} gauge theory.  Tensor networks \cite{weichselbaum2012non,Orus:2013kga,Fishman:2020gel,Meurice:2020pxc} and the Density Matrix Renormalization Group (DMRG) are 
then used to compute the mass spectrum.  
We find that QZD exhibits a fascinating and unexpected relation between the masses of the lightest fermions and bosons.

All states measured are gauge invariant, and so confined.
This is encouraging, as there is a long history suggesting
that confinement in both $2+1$ and $3+1$ dimensions are
dominated by the \Zthree vortices of \SU{3} gauge theories
\cite{Greensite:2016pfc,Sale:2022qfn,Biddle:2022zgw,Biddle:2022acd,Biddle:2023lod}.  In $1+1$ dimensions, these \Zthree vortices are points in spacetime, but should also confine.

\section{QZD and its weak and strong coupling limit}
Our starting point will be the following standard lattice Hamiltonian 
\begin{eqnarray}
  H_L \!&=&\! -\tfrac{i}{2a} \sum_{x=1}^{L-1}
    \bigl( U_x^\dagger\, \bchi^\dag_{x} \cdot \bchi_{x+1}
     {-} \mathrm{H.c.}\bigr)
    - m \sum_{x=1}^L  (-1)^x {n_x}
   \notag \\
   &&+ \tfrac{ag^2}{2} \sum_{x=1}^{L-1} E_x^2
\text{ ,}
\label{eq:discH}
\end{eqnarray}
where $\bchi_x {\equiv}
(\chi_1,\ldots,\chi_{N_f})_x^T$ are staggered
fermions of $N_f$ flavors that live
on even/odd sites representing the original left/right chiralities.
The particle number at a site
\begin{equation}
   n_x \equiv \bchi_x^\dagger \cdot \bchi_{x} \equiv 
   \sum_{f=1}^{N_f} \chi_x^{f\dagger} \chi_{x'}^f
\text{ ,}\label{eq:nx}
\end{equation}
includes a symmetric sum over all flavors.
It follows from \Eq{discH}, that like $E_x$, $U_x$ 
lives on the bond in between sites $x$ and $x+1$.
We consider a finite system with a total of $L$ sites
together with open boundary conditions (BC).
The unit of energy is assumed 
in terms of the hopping amplitude $1/(2a):=1$,
i.e., $a=1/2$, unless specified otherwise.
For the remainder of the paper, we focus on the case
of $N_f=3$ fermionic flavors.

The model differs from the Schwinger model in that $U_x$, $E_x$
and Gauss law  implement a local \Zthree algebra
\cite{Horn:1979fy,Kogut:1980fn,Kogut:1980qb,Alcaraz:1980bb,Alcaraz:1980sa,Zohar:2016iic,Ercolessi:2017jbi,Magnifico:2018wek,Magnifico:2019ulp,Borla:2019chl,Frank:2019jzv,Emonts:2020drm,Robaina:2020aqh,Emonts:2022yom}.
Defining the operator
\begin{eqnarray}
   P_x &\equiv& \exp\left( \tfrac{2\pi i}{3} \;  E_x\right)
\end{eqnarray}
we impose 
\begin{eqnarray} 
   && P_x^3 = U_x^3 = 1 \ ; \quad P_x^\dagger
      P_x^{\,} = U_x^\dagger U_x^{\,} = 1 
\\ && U_x P_x = {\rm e}^{2 \pi i/3} P_x U_x
\ .  
\end{eqnarray}
In the basis where the electric field is diagonal,
$U_x$ takes the role of a cyclic permutation operator,
\begin{equation}
    U_x = \begin{pmatrix} 
    0 & 1 & 0\\
    0 & 0 & 1 \\
    1 & 0 & 0
    \end{pmatrix} \ ,
\end{equation}
that increments (or for $U^\dagger$ decrements
the gauge field.
This is supplemented by a \Zthree Gauss law
\begin{equation}
   P_{{x}}^{\,} P_{{x-1}}^\dagger
 = \exp\left( \tfrac{2\pi i}{3} \, q_x \right) \ .
\end{equation}
with the charge density defined as usual for staggered fermions

\begin{eqnarray}
  q_x = \left\{
     \begin{array}[c]{ll}
       n_x & \text{for } x \text{ odd}\\
       n_x - N_f & \text{for } x \text{ even } \ (N_f=3) \ .
     \end{array}
  \right.
\label{eq:qx}
\end{eqnarray}
This permits the simple interpretation
that odd sites behave like `particles'
which carry electrical charge $+1$, thus having $q_x=(+1)\,n_x$,
whereas even sites behave like `holes',
carrying electrical charge $-1$ for every hole
relative to completely filled, thus having
$q_x = (-1)\, (N_f - n_x) = n_x - N_f$.

While the variables are similar to the implementation of a \U{1}
gauge theory by quantum links
\cite{Chandrasekharan:1996ih,Brower:1997ha}, Gauss' law is
different, as the flux is only conserved modulo 3.  
We further massage Eq. \eqref{eq:discH} to make it more amenable
to numerical simulations. We start by imposing open boundary
conditions on our chain $E_0=E_{L}=\chi_{L+1}=0$. This allows us
to use the remaining gauge transformations to remove the links
$U_x$ from the theory (see for instance
Ref. \cite{Chakraborty:2020uhf}) and solve Gauss' law, expressing the
electric field operators in terms of the fermionic fields.  We
have
\begin{align}
    E_{x} &= \left( Q_x \,{\rm mod}\, 3 \right) 
\label{eq:Ex}
\end{align}
with the cumulative charge
\begin{align}
    Q_x &\equiv \sum_{{x'\leq x}} {q_{x'}} \ ,
\label{eq:Qx}
\end{align}
and where modulo is taken symmetric around zero, i.e.,
having $E_x \in \{-1,0,1\}$.
Thus in $1+1$ dimensions the gauge fields are not dynamical,
as they can be completely determined by the charge
configuration. This permits one to express
a long-range Hamiltonian
entirely in terms of the fermion fields.
By exploiting Abelian \U{1} particle number symmetry
in the simulation, this is conveniently done relative
to half-filling all along. With this then
the symmetry label for the cumulative block particle number
\begin{align}
    N_x &\equiv \sum_{{x'\leq x}} ({n_{x'}-n_0}) \ ,
\label{eq:Nx}
\end{align}
with $n_0 = N_f/2$ the average half-filling
directly specifies $Q_x$ for even block size,
i.e., $Q_x = N_x$.
For odd block size this requires a minor tweak
based on \Eq{qx} ensuring that $Q_x \in \mathbb{Z}$.

A continuum form of a \Zthree gauge theory can be constructed following Krauss, Preskill, and Wilczek \cite{Krauss:1988zc,Preskill:1990bm,Pisarski:2021aoz}.  
One begins with a \U{1} gauge field, coupled to fermions with unit
charge, and a scalar field, $\phi$, not with unit charge, but with charge {\it three}.  Arranging
the potential for the scalar field to develop an expectation value in vacuum, $\phi_0$, the photon develops a mass $m_\gamma = 3 g \phi_0$,
and so is screened over distances $> 1/m_\gamma$.
Since the scalar field has charge three, the $\phi$ field is insensitive to the presence of \Zthree vortices, which leaves a local \Zthree symmetry, at least over distances $> 1/m_\gamma$.  Remember that a scalar field has zero mass dimension in two dimensions, so by taking $\phi_0 \gg 1$, the \U{1} photon is very heavy, and the theory only goes from the effective \Zthree gauge symmetry, to the full \U{1}, at short distances $\leq 1/m_\gamma$.

\subsubsection*{Symmetries}

The states in the theory can be labeled by their total particle
number $N_{\rm tot}$ which we take relative to half-filling
for convenience,
and the representation of \SUf{3} flavor
symmetry $(n_s,n_a)$ to which they belong. Here $n_s$/$n_a$
denote the symmetric/antisymmetric rank of the representation.
We then use $(N_{\rm tot}; n_s n_a)$ as a compact notation to 
label all symmetry sectors,
as explained in \App{symlabels}.  We restrict
ourselves to the ground state sector $(0;00)$,
and the lightest states $(0;11), (3;00), (3;11)$.
Here $(11) \equiv {\bf 8}$ (octet) specifies the adjoint
representation of \SU{3}.  Mesons live in the $N_{\rm tot}=0$
sector, while baryons live in the $N_{\rm tot}=3$ sector.  As explained in the Introduction, this is unlike a
\U{1} theory,  \textit{since both sectors can be
realized in the absence of any net external charge},  i.e. in
the gauge invariant sector which satisfies Gauss's law. The ground state
in the $(0;00)$ sector represents the QZD vacuum,
with no baryon or meson excitations present.

\section{Results}

\subsection{Weak coupling regime}

Naively, one might expect that the weak coupling behavior of this theory would be the usual power series in $g^2$.  To understand why this is not so, start first with the case of a \U{1} gauge theory in two spacetime dimensions.  In the continuum,
the Coulomb potential is
\begin{equation}
 V(x)= g^2 \int dk \, \frac{{\rm e}^{i k x}}{k^2} \sim g^2|x| \,
\end{equation}
is confining.  For very small coupling, the fermions are heavy, and we should be able to use a non-relativistic approximation:
\begin{equation}
{\cal H}_{\rm non-rel} = - \frac{1}{2 m} \frac{d^2}{d x^2} + g^2 \frac{|x|}{4}
\, \; .
\end{equation}
Because this is a confining potential, the weak coupling expansion is not a power series in $g^2/m^2$, but in
$(g^2/m^2)^{1/3}$ \cite{Fonseca:2006au,Fateev:2009jf,Ziyatdinov:2010vg,Zubov:2015ura}. In App.~\ref{App:weakcoupling} we show that the meson mass behaves as
\begin{equation}
\frac{M_{\rm meson}}{m} = 2m \cdot 
\left(
1 + 0.40431  \cdot \left(\frac{g}{2m}\right)^{4/3} + 
O\left(\frac{g^2}{m^2}\right) 
\right)
\label{eq:weakcouplingmain}
\end{equation}

\begin{figure}[]
\centering
\includegraphics{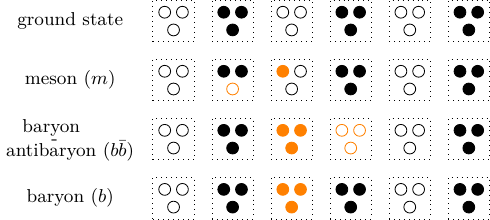}
\caption{
    Illustration of the strong coupling limit.
    The ground state (top row) corresponds to half-filling. 
    The first excitation is a single baryon (last row).
    We highlight in orange the deviation from the ground state.
    The first ``mesonic''  state is a baryon-antibaryon pair
    which in QZD does not generate an electric field (by contrast,
    a single meson requires partial fillings; since necessarily also
    delocalized, by Gauss law this will always generate electric field,
    and hence decouple at strong coupling to large energies).
}
\label{fig:strongcouplingvac}
\end{figure}

\subsection{Strong coupling regime}
\label{sec:strong_coupling}

Another limit that is under control is the strong coupling region of the lattice model, keeping the lattice spacing $a$ fixed as $g\to\infty$).
The vacuum at infinite coupling is elementary, and direct to
expand about. 
In terms of spins, it corresponds to half-filling: 
for each flavor, all even sites are occupied, while all odd
sites are empty, as in Fig.~\ref{fig:strongcouplingvac}. The first excitation is a ``baryon", with one fermion of each flavor sitting at the same site. Thanks to the periodicity of Gauss's law for \Zthree, such a configuration has zero net charge.
Thus to zeroth order in $1/g$, the mass of the baryon 
is just $3m$, see (\Fig{fullspectrum} below).
The leading correction in $1/g^2$ comes from the virtual hopping of a single fermion.
This hopping costs $ag^2$ in energy, and occurs
with probability $1/(4a^2)$, in $3L$ possible ways.
To leading order in perturbation theory, the baryon
mass is then shifted by
\begin{equation}
  m_B = 3 m + 
        3 L \cdot \frac{1}{4a^2} \cdot \frac{1}{ag^2}
      = 3m + \frac{3}{4 a^3}\frac{1}{g^2} \ .
\label{baryon_mass_strong}
\end{equation}

In contrast, mesons behave very differently in strong coupling. 
 Consider first a meson in the adjoint representation.  To carry net flavor, they must be composed of a fermion on one site and an anti-fermion on an adjacent site, so
 unavoidably there is a nonzero electric flux connecting the two.
 As the energy from a single link is $\sim g^2$, 
 adjoint mesons are very heavy at strong coupling,
 with a mass $\sim g^2$.  Further, at $g^2 = \infty$ they are small, only a single link in size.
 
Somewhat unexpectedly, this is {\it not} true for a meson which is a flavor singlet.  For a \Zthree gauge theory,
three fermions of different flavors, $uds$, are themselves a singlet under \Zthree.  Thus at infinite coupling, we can form a singlet meson by putting $uds$ on one site, and 
$\overline{u}\, \overline{d}\,\overline{s}$ on {\it any} other site --- no matter how far apart!  At $g^2 = \infty$, then, the
mass of the flavor singlet meson is just $6m$.

For large but finite coupling, the positions of the 
$uds$ and $\overline{u}\, \overline{d}\, \overline{s}$ are
correlated with one another, as the singlet meson
mixes with three adjoint mesons.  To $\sim 1/g^2$ one can show that the correction to the mass of the  singlet meson is
identical to that of the baryon, Eq.
(\ref{baryon_mass_strong}).  

The size of the singlet meson is also surprising.
At infinite coupling it is of {\it infinite} size, with
the size of the singlet meson large when $g^2$ is large.

\subsection{DMRG spectra}
\label{sec:DMRG}

In order to access the spectrum at all couplings, we perform
simulations using the Density Matrix Renormalization Group (DMRG). We take
full advantage of the flavor \SUf{3} global symmetry of our
system by using the \mbox{QSpace} tensor network
library \cite{Wb12_SUN}, which is highly efficient. 
Utilizing this symmetry also
allows us to target different symmetry sectors and gives us
direct access to lowest lying excitations.  

\begin{figure}
\includegraphics[width=1\linewidth]{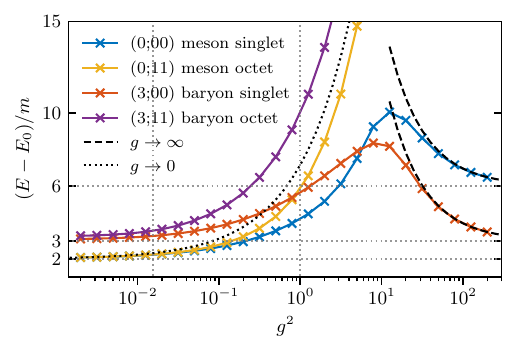}
\caption{
   Low lying excitation energies vs. coupling $g^2$ for
   mass $m=0.125$ ($L=120$). The red and blue lines correspond
   to the lowest lying \SUf{3}-singlet baryons (red) and
   mesons(blue). Their behavior is qualitatively different from
   the lowest \SUf{3}-octet baryon (purple) and mesons (yellow),
   since because of the \Zthree gauge symmetry, the singlets do not decouple in the limit of strong coupling. 
   The horizontal guides (black dashed) indicate limiting
   values for weak and strong coupling.
   The vertical guides indicate the value $g=m$
   which separates weak from strong coupling,
   as well as $g=1/2a=1$ where the interaction becomes equal
   to the hopping amplitude (the continuum limit only
   has access to $g\lesssim 1$).
}
\label{fig:fullspectrum}
\end{figure}

We show the spectrum in Fig.~\ref{fig:fullspectrum}, for a given
mass $m$, as a function of $g^2$. We show the energy difference
between the lowest lying state above the vacuum in a given
symmetry sector and the vacuum, normalized by the bare mass $m$.
The particle content of these states can be easily identified in
two limits. At weak coupling, the singlet and octet states are
degenerate.
For $N_{\rm tot}=0$, they correspond to a single meson
of mass $\approx 2m$.  For $Q_{tot}=3$, they correspond to a
single baryon of mass  $\approx 3 m$. While for a baryon
we can put $uds$ on a single site and satisfy
Gauss's law for the gauge group,
we cannot do this for mesons.
In weak coupling mesons are created by putting a fermion
on one site, and an anti-fermion on another site.
This implies that they creat a nonzero value for
\Zthree electric flux. This does not
matter at weak coupling as 
contributions to the energy from electric flux is small,
a fractional power of $\sim g^2$. 

Given the discussion above, the particle content of these states is easy to identify in weak and strong coupling.
A meson with symmetry $(0;00)$ continuously interpolates from a single meson at weak coupling to the $b\bar b$ excitation at large coupling. 

At weak coupling, the singlet and octet states are
degenerate, with mass $2m$ at $g^2 = 0$.
For $N_{tot}=3$, there is a single baryon whose
mass is $3 m$ at $g^2 = 0$. 

\begin{figure}
\includegraphics{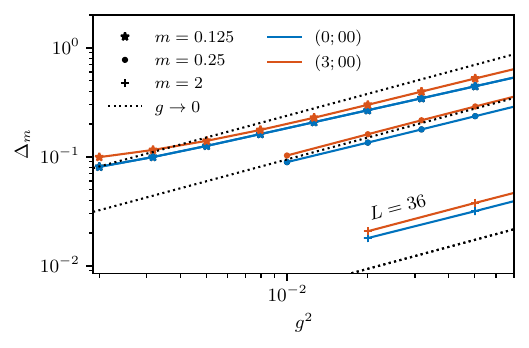} 
\caption{
Weak coupling regime ($g^2\ll1$)
--
Dependence of the gap on the coupling constant after subtracting the free case value $\Delta_m= (E-E_0-E_{\rm free})/m $ with $E_{\rm free}=2m,3m$ for mesons and baryons, respectively.  We show the meson (empty purple) and baryon (filled gray) values for different masses. The dotted lines correspond to the weak coupling  $(g^2)^{2/3}$  expansion of \cite{Ziyatdinov:2010vg}. The heavier mass is deep in the lattice regime and unsurprisingly shows large deviation. For smaller masses, the $\mathrm{QCD}_2$ is surprisingly close to the \Zthree data. We further discuss the remaining finite volume effects in Fig.\ref{fig:weakcoupling:2}. The $(g^2)^{2/3}$ is a striking indication of confinement.
}
\label{fig:weakcoupling}
\end{figure}

The behavior of the masses as the coupling constant increases
is shown in Fig. \ref{fig:weakcoupling}. 
It is striking that the mass
of the adjoint meson agrees well with the perturbative
result of Eq. \eqref{eq:weakcouplingmain}, which we
compute only up to leading order, up to rather large coupling,
certainly up to $g \sim 1$.
In contrast, by $g \sim 1$ the result for the singlet meson is significantly lower than the perturbative result at leading order.  This is natural because the singlet meson of QZD
has no analogy in either the 't Hooft model or in QED.

At strong coupling,
the first excited state in the $(0;00)$ channel
corresponds to multiparticle states, including both the baryon-antibaryon ($b\bar{b}$)
and states with three mesons.
The dotted lines show the
leading $1/g^2$ corrections, Eq. \ref{baryon_mass_strong}. 
 There is good agreement with our numerical data. 

In particular, the fact that the octet meson becomes
heavy in strong coupling, and that 
the $(0;00)$ sector are heavier than the $(3;00)$ sector at
strong coupling, indicates that there are two values of the
coupling constant where there is a degeneracy between
a baryon and meson state.
As the coupling increases, the first is where the singlet baryon is degenerate with the octet meson. The second, at larger coupling, is where the
singlet baryon and the singlet meson are degenerate. Note that
this prediction is specific to \Zthree, as even the singlets
decouple in \U{1}. This is illustrated in 
Fig.~\ref{fig:fullspectrum}.
These two crossings may simply be fortuitous.
The second crossing, where the singlet baryon and singlet
meson are degenerate, is suggestive of supersymmetry.
However, we have not checked whether this degeneracy remains
true for the excited states at higher mass.

The precision of our data also allows us to confirm that the theory confines. In particular, we can extract the small coupling dependence of the  mass gap. We illustrate this in Fig.~\ref{fig:weakcoupling}. We plot the energy relative to the ground state minus the $g=0$ contribution in the continuum, $\Delta_m= (E-E_0-E_{free})/m $ with $E_{free}=2m$ and $3m$ for mesons and baryons, respectively. A confining potential leads to non-analycities in the coupling strength and as argued above, in an expansion in $g^{4/3}$ instead of $g^2$. We show data for different masses for the singlet meson and baryons. We also show the prediction of \cite{Ziyatdinov:2010vg} with dotted lines. For larger masses, we see strong deviations, which is completely expected as this is deeply in the lattice regime of our model $m\sim 1/a$. For smaller masses, the prediction agrees well with our data. Deviations from the $(g^2)^2/3$ at small $g^2$, most prominent for mesons, can be attributed to finite-size effects. 

\begin{figure}
\includegraphics{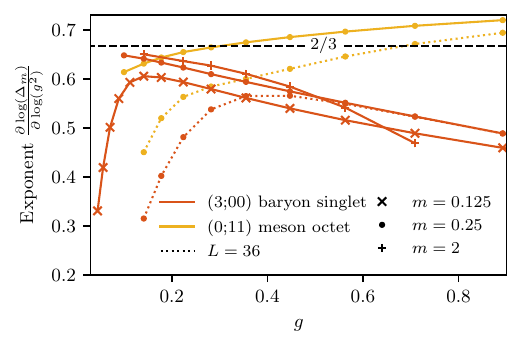}
\caption{
Leading exponent for small $g$ --
Logarithmic derivative of data in \Fig{weakcoupling}.
All the slopes converge to an exponent of $(g^2)^{2/3}$. The bending of the curves away from the limiting value at small $g^2$ signal finite volume effects in this region. This is best appreciated by comparing the dotted circle with the plain circle; they correspond to data at the same mass but different volumes.
}
\label{fig:weakcoupling:2}
\end{figure}

We make this more quantitative In Fig.~\ref{fig:weakcoupling:2}
We estimate the exponent of $g$ by computing the logarithmic derivative of $\log(\Delta_m)$, which gives an estimate of the leading exponent at small $g^2$. We first show the result for the baryon singlet (dark orange lines). The exponent converges to $(g^2)^{2/3}$ for all different masses. We can also cleanly identify finite-size effects: they bend the curves away to zero, as seen by comparing the data at $m=0.25$. The plain line corresponds to $L=120$ while the dotted one to $L=36$. By looking at the same quantity for the mesons, we can substantiate our claims that the finite volume effects are stronger in the sector, consistent with a smaller gap. We show in yellow the behavior of the meson octet at $m=0.25$ for $L=36$ and $L=120$. The exponent still shows a strong dependence on volume size. The trend is however consistent with the $(g^2)^{2/3}$ expectation, confirmed in the baryon channel.
 

\subsection{{Topological edge modes vs. bulk excitations}}

Beyond the spectrum, we
also study the spatial distribution
of excited states. Because of staggering and open boundary
conditions, the spatial structure of the ground state is
non-trivial.  Indeed, the use of staggered fermions in the
Hamiltonian \eqref{eq:discH} gives it a simple topological
nature with topologically protected edge modes
at the open boundaries for the ground state.
We emphasize, though, that this already also holds for the plain
non-interacting model in the absence of any gauging, i.e.,
$g=0$, in which case the topological aspect is known as the
Su-Schrieffer-Heeger (SSH) model \cite{Su79,Batra20}.
However, at finite $g$ this raises several 
non-trivial questions:
(i) do the edge modes remain topologically protected
when turning on finite $g$?
(ii) if yes, how, are these edge modes characterized
in terms of excess particle number and excess electric charge?
(iii) to what extent is the nature of the
excited states affected by the presence of open boundaries,
i.e., are the excited states true bulk modes,
or rather a property of the boundary?

\begin{figure}[tb]
\begin{center}
\includegraphics[width=1\linewidth]{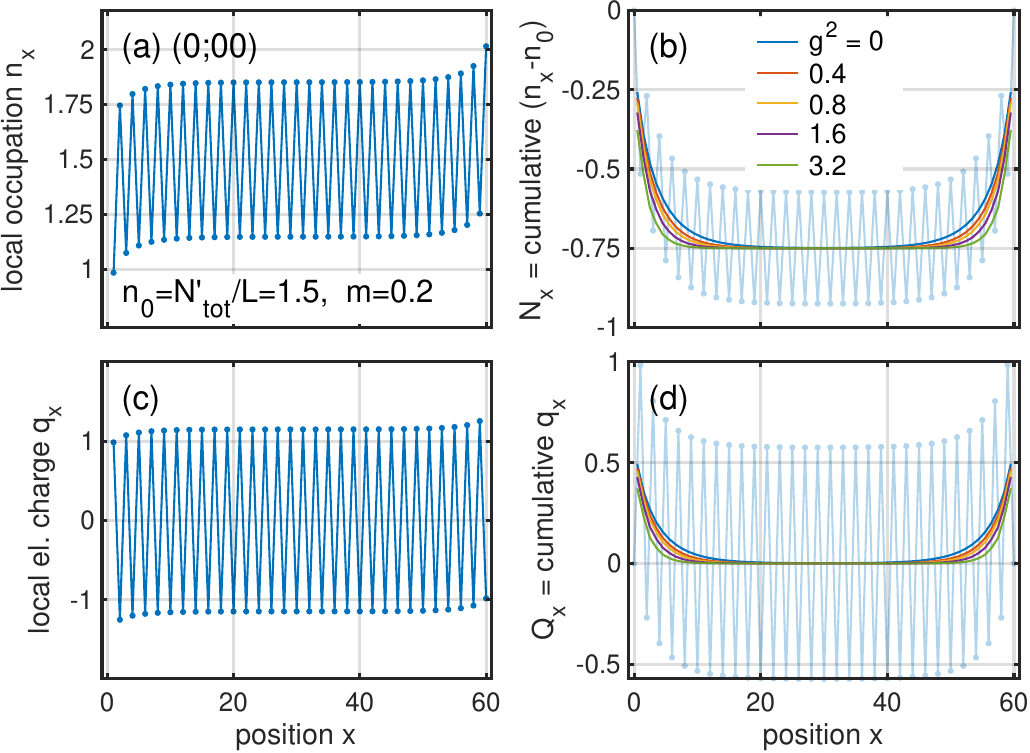}
\end{center}
\caption{
  Edge modes for the QZD ground state,
  i.e., in the symmetry sector (0;00) for a system of
  size $L=60$ for $m=0.2$ -- 
  (a) Local particle number
  $n_x \equiv \langle\bchi_x^\dagger \cdot \bchi_{x}\rangle$
   vs. position $i$ along the chain, for the free model $g^2=0$.
  Here $N'_{\rm tot} \equiv \sum_x n_x$ is the actual,
  number of the filled Fermi sea given a finite lattice size.
  (b) Cumulative data in (a), after subtracting
  half-filling $n_0 = N_f/2 = 3/2$ for each site,
  i.e., plotting $N_x \equiv \sum_{i'=1}^{i} (n_x - n_0)$.
  This data (light blue) still shows alternating behavior.
  Averaging over even and odd $x$ yields the smooth curve
  (solid darker blue).
  Additional averaged data for finite $g$ is presented 
  in different colors, where the respective value of $g^2$
  is specified in the legend.
  (c) Local electric charge based on \Eq{qx}.
  (d) (Averaged) cumulative data of (c)
  [similar analysis as in (b), also sharing the same legend].
}\label{fig:edge000} 
\end{figure}

The edge mode in the ground state for the non-interacting case
($g=0$) is analyzed in \Fig{edge000}(a) at mass $m=0.2$ for
an $L=60$ system.  Clearly, the alternating onsite energy
$\varepsilon_x = m(-1)^{x-1}$ directly translates to even/odd
variations of the local occupations around the average filling
$n_0=3/2$ (half-filling) throughout the system. However:
this occupation pattern changes systematically towards the open
boundaries.  The data in \Fig{edge000}(a) bends down at left
boundary, and up at the right. The cumulative local particle
number relative to half-filling, $N_x \equiv \sum_{x'=1}^{x}
(n_x - n_0)$, is shown in \Fig{edge000}(b), in light blue in
the background.  As this data is still alternating
around a well-defind mean value, it is averaged over
even and odd lengths [darker blue for $g=0$ in
\Fig{edge000}(b)]. This averaged data $\bar{N}_x$ shows
that the particle number offset due to the open boundary is
$n_{\rm edge} = \bar{N}_{x=L/2} = 3/4$.
The precise nature of the averaging matters here:
by the procedure above,
$\bar{N}_x = \sum_{x'=1}^{x-1} (n_{x'} - n_0) + (n_x-n_0)/2$.
If instead, for example, one had computed the
cumulative particle number over unit cells which pairs
up neighboring sites, the resulting excess particle
number would not have been strictly universal.

Eventually, the cumulative excess particle number on the left
boundary is exactly compensated at the right boundary. The
cumulative total particle number offset over the entire system
again returns to zero in \Fig{edge000}(b).  Therefore the
excess particle number of the edge modes have the same value,
but opposite signs for the two boundaries.

A non-zero interaction $g$  increases the gap in the system, see
\Fig{fullspectrum}. Consistently, the edge modes localize
more towards each open boundary (other colored lines in
\Fig{edge000}(b)).  The topological aspect of the
non-interacting model remains  preserved as long as the gap does
not close. Conversely, the topological protection remains intact
in the presence of finite gauge strength $g$.

The value of the fractional excess particle number
can be motivated straightforwardly for $g\to \infty$:
there one has a simple product state of alternating completely
empty and filled sites [see first line (ground state) in
\Fig{strongcouplingvac}].
Therefore starting from the left open boundary, 
the particle number relative to half-filling is given by
$n_x-n_0 = [-n_0,+n_0, -n_0,+n_0, \ldots]$ with $n_0=3/2$.
Its cumulative sum
is $[-n_0,0, -n_0,0, \ldots]$. This averages to $-n_0/2$,
and therefore $n_{\rm edge} = 3/4$.
This is precisely the excess number of particles
observed in \Fig{edge000}. For the extremal case
here this excess particle number is strictly located right at
the boundary.
When reducing $g$, the edge mode starts to reach into the
system as seen with \Fig{edge000}(a). 
The cumulative excess particle number with each open
boundary, nevertheless, remains pinned to precisely the same value 
\begin{equation}
   n_{\rm edge} = \frac{3}{4} = \frac{N_f}{4} 
\text{ .}\label{eq:nedge}
\end{equation}
By having an odd number of flavors here, this shows
that the edge mode carries a fractional particle number.
This persists for any value of $g$ all the way down
to $g=0$ since the gap of the system never closes. 
Hence as long as the system is long enough, such that the
overlap of the tails of the boundary modes is negligible in the
system center, one always obtains precisely the same value
$\pm n_{\rm edge}$ for the excess particle number
with opposite sign for the left and right boundary.
Since this includes $g=0$, this shows that
the topological protection of the SSH model
remains intact also when gauging the system.
Indeed, what protects SSH is inversion symmetry \cite{Wang23}.
Gauging leaves this symmetry intact, e.g., 
for infinite systems
or periodic systems of even length.

Now for a lattice gauge theory, by having an excess number
of particles associated with an edge, one may worry
that there is an electric field throughout the bulk
connecting the two excess particle numbers of opposite
sign for each boundary. However, this is not the case:
while there is an excess number of particles due to
the edge mode in the ground state, it {\it does not} carry
any net effective electrical charge, therefore $q_{\rm edge} = 0$.

This is demonstrated in the lower panels of \Fig{edge000}
which repeats the same analysis as in the upper panels,
but now for the electrical charge, using the number to charge
conversion in \Eq{qx}.
From the analysis in panel (d) one finds
$q_{\rm edge} = \bar{Q}_{x=L/2} = 0$.
The smooth averaged curves in \Fig{edge000}(b)
simply got shifted to the zero base line in \Fig{edge000}(d).
This can be similarly motivated as for excess
particle number above for the case $g\to\infty$: 
given the product state with alternating
completely empty and filled sites, 
in the present case one obtains for the charge,
starting from the left boundary, $q_x=[0,0,0,\ldots]$
which averages to zero, indeed.

\begin{figure}[tb]
\begin{center}
\includegraphics[width=1\linewidth]{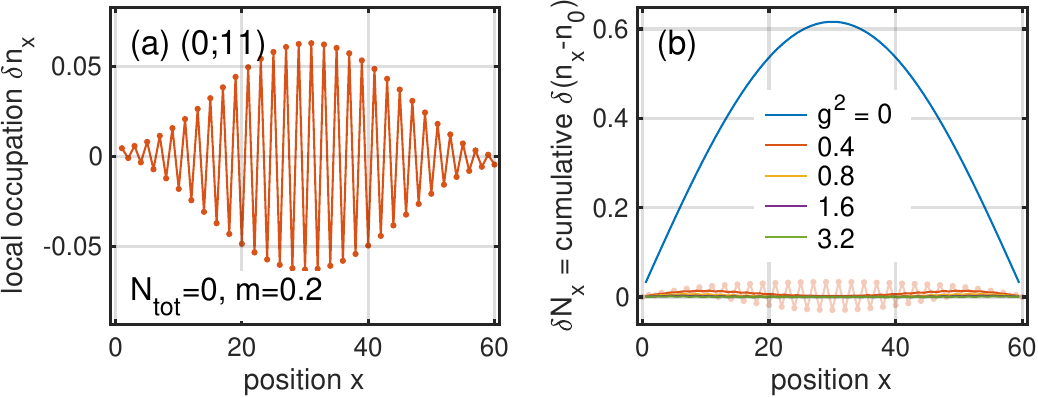}
\end{center}
\caption{
   Lowest octet meson excitation
   [lowest energy eigenstate in the (0;11) symmetry sector] --
   (a) Difference of local particle number $\delta n_x$
   relative to the ground state for $g^2=0.4$
   [right legend also applies to panel (a);
   same parameters as in \Fig{edge000} otherwise]. 
   (b) Cumulative data of (a) starting from the left boundary
   (light red) which is again even/odd averaged (solid red).
   Other smooth lines are obtained the same way for different
   $g^2$ as specified with the legend.
   Since this is a meson, eventually $\delta N_{\rm tot}=0$.
}\label{fig:edge011} 
\begin{center}
\includegraphics[width=1\linewidth]{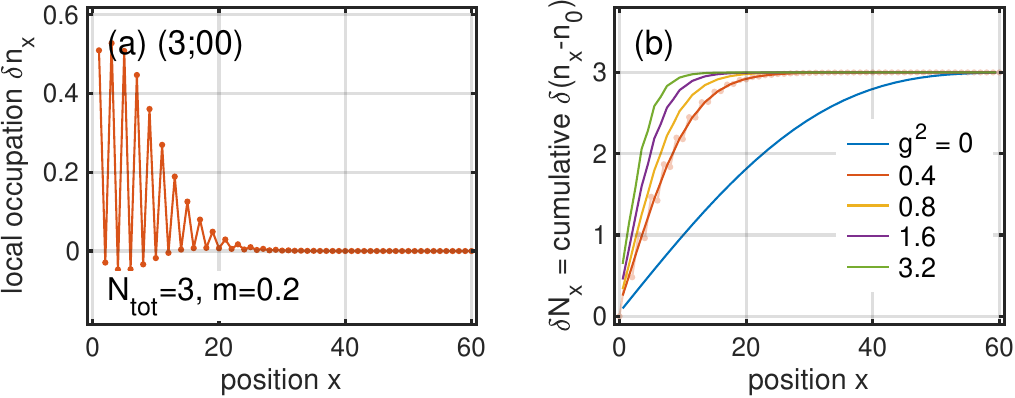}
\end{center}
\caption{
   Lowest baryon excitation 
   [lowest energy eigenstate in the (3;00) symmetry sector] --
   same analysis as in \Fig{edge011} otherwise.
   Since this is a baryon, eventually $\delta N_{\rm tot}=3$.
}\label{fig:edge300} 
\end{figure} 

\begin{figure}[tb]
\begin{center}
\includegraphics[width=0.85\linewidth]{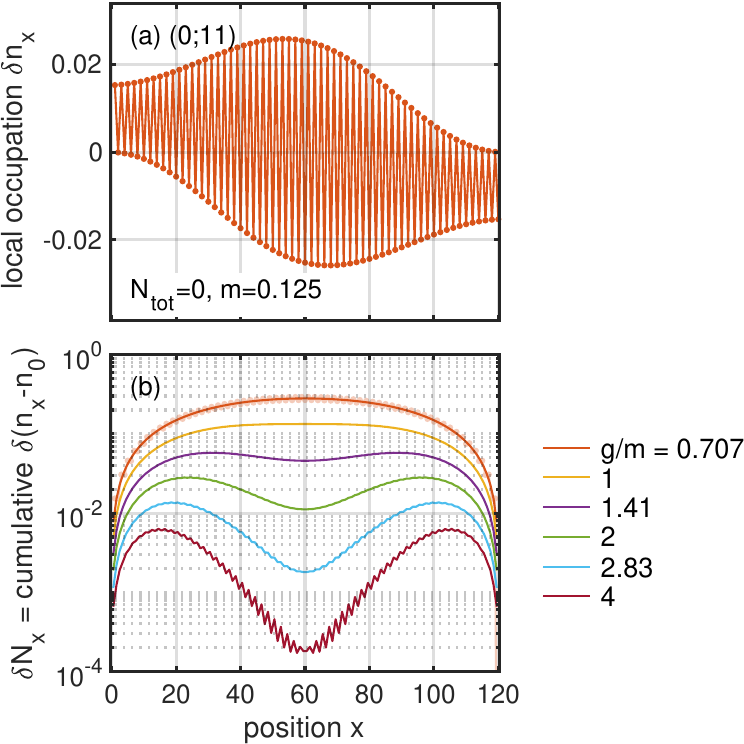}
\end{center}
\caption{
   Same analysis as in \Fig{edge011}, yet for the system
   parameters as in \Fig{fullspectrum}, i.e., for twice
   the system size $L=120$ here, while at the same time
   also smaller values of $g^2$ are used. Having $m=0.125$,
   the legend thus implies $g^2 \leq (4m)^2 = 0.25$.
   The $x$-axis in (a) is the same as in (b),
   with (b) shown on semilog-y scale as compared to 
   \Fig{edge011}, in order to focus on the splitting
   of the data into a double peak structure.
}\label{fig:edge011c} 
\end{figure}

Having a clear understanding of the
edge modes due to the open boundaries
as discussed in \Fig{edge000}, we now turn to
excited states. Specifically, we want to ensure that low-energy
baryon or meson excitations are true bulk excitations, and not a
consequence of the presence of the open boundaries.
In \Fig{edge011}(a) we show the spatial distribution of the
differential particle number occupation $\delta n_x$ for the 
octet meson $(0;11)$ relative to the ground state
for $g^2=0.4$ (same parameters as in \Fig{edge000}).
The variations throughout the entire system clearly
demonstrate the bulk nature of this excitation.
The cumulative sum of the variation in \Fig{edge011}(a)
is shown in \Fig{edge011}(b), supporting a similar
picture. Since the total filling remained the same
as for the ground state, the data in \Fig{edge011}(b)
returns to $\delta N = 0$ for $x=L$.
The variations in \Fig{edge011}(b) diminish quickly, though,
when increasing $g^2$ (smaller $g^2$ values will be 
analyzed in \Fig{edge011c}).

The lowest singlet baryon (b) excitation [(3;00) symmetry sector]
is analyzed in \Fig{edge300}. Analogous to the
meson flavor excitation in \Fig{edge011},
this again plots the differential variation of the particle
number occupations $\delta n_x$ relative to the ground state.
\FIG{edge011}(a) suggests that the baryon is 
(weakly) attracted to the left boundary.
It is still a bulk excitation, though, in the sense that its
extent clearly exceeds the penetration depth of the
edge mode for the same $g^2=0.4$ as compared to \Fig{edge000}(b).

By adding a baryon to the system, it is free to propagate.
Via the kinetic term in the Hamiltonian (hopping term),
the baryon has a tendency to delocalize across the entire
system. Because of the gauge field, however, this motion
generates electric fields
which cost energy. Therefore in the presence
of open boundaries, this energy is minimized
by putting some of the excess particle number of 
$\Delta N_{\rm tot} = 3$ right at the very first site of
the left boundary as this site is particle-type:
being below half-filled, this can hold more extra particles.
Since there is no hopping to the left of the first site,
there is less energy cost in terms of the
electric field this would generate. This weak energetic bias
towards the left boundary therefore is related to the 
convention that the system starts with particle-like
site, i.e., with local energy $\varepsilon_1 = m (-1)^0 > 0$.
For this reason, we expect an isolated antibaryon ($\bar{b}$)
to be attracted to the opposite boundary at the right.
From this perspective, one may expect that
the meson in \Fig{edge011}(b) for sufficiently strong $g^2$
starts to split a $b\bar{b}$ pair separated to
opposite boundaries. This is supported by the weak
double peak structure that develops in \Fig{edge011}(b)
for larger $g^2$, indeed. Clear evidence for the same
will be provided in \Fig{edge011c}. 

It is instructive
to track how excitations are distributed over a finite
system with open boundaries as the interaction $g$ is increased.
Let us start by discussing the baryon excitations.
As one would expect from the continuum theory, the larger the coupling strength, the more localized the baryon state 
can become around a perturbation of an otherwise
uniform system. In the present case this perturbation
is given by the abrupt end of the system
due to the open boundary.
We expect such localization also to carry over to
the lattice model. In the extremal case $g\to\infty$ where the ground
state is a simple alternating product state as
depicted at the top of \Fig{strongcouplingvac},
the baryon excitation simply fills any of the
particle-like sites (last row in \Fig{strongcouplingvac}).
This results in degeneracy, and thus a flat-band excitation.
For large but finite $g$, there is a weak preference
on the first site (left boundary) because of the
earlier argument. This more apparently still here for large $g$,
since the QZD interaction far dominates the kinetic energy.
From the QZD perspective, due to the \Zthree setup
an excess charge of $\delta N_x = 3$ does not
generate an electric field since in that case
the electric charge is effectively zero, $Q_{\rm tot}=0$.
Hence for larger $g$, this confines the baryon
in the neighborhood of the boundary
as is seen in \Fig{edge300}(b):
the data quickly transitions from $\delta N_x = 0 \to 3$.
Both, excess particle and electric charge are attracted
to the left boundary. 
For $g\to 0$ eventually, the bias of the above
type diminishes. At $g=0$ the baryon
excitation is a true bulk excitation that is symmetric
around the system center [blue line in \Fig{edge300}(b)]
up to even/odd alternations.

In this light we return to the low-energy mesons.
We argued with the spectra in \Fig{fullspectrum}
that the meson singlet [in (0;00)] starts from a single
particle/hole pair at weak coupling.
For strong coupling, however, this becomes a $b\bar{b}$ pair.
In \Fig{strongcouplingvac} (third line) this is exemplified
locally by shifting the particles from a
completely filled site to a neighboring completely
empty site.

From the present analysis we find that a single baryon
is attracted to the left open boundary. By symmetry
we argued that the antibaryon is attracted to the opposite
boundary. Hence in the presence of $b\bar{b}$ meson
at strong coupling, we expect due to the presence
of the open boundaries, that the $b\bar{b}$ pair
is dissocated towards the open boundaries
as this permits a weak energy gain.

Revisiting \Fig{edge011}(b), we find, indeed, that for larger
$g^2$ a weak double peak structure develops in the data. In
order to focus on this behavior, we repeat the analysis in
\Fig{edge011}(b) for the system parameters in
\Fig{fullspectrum} in \Fig{edge011c} (hence twice the system
length, yet also smaller $g$ values). By specifying $g$ in units of
$m$ in the legend of \Fig{edge011c}(b), we find that the double
peak structure develops around $g\simeq m$. At close inspection,
the same also holds for the parameters in \Fig{edge011}.

Hence the appearance and dissocation of the $b\bar{b}$ meson
occurs far before the peak in the data towards large $g$ in
\Fig{fullspectrum}: that peak in \Fig{fullspectrum}
is located around $g \sim 1/a$ where the coupling $g$ becomes
stronger than the one-particle bandwidth. While the latter is a
pure discretization effect, the dissocation of the $b\bar{b}$
occurs much sooner around $g\sim m$. Hence this behavior is
expected to be a true property of QZD also in the continuum
limit. The transition towards a $b\bar{b}$ meson around $g\sim
m$ thus is consistent with the intuitiv notion that $g\sim m$
separates the weak from the strong coupling regime in the
lattice gauge theory.

In the weak to intermediate coupling regime, the ground state
(QZD vacuum) is far from the plain product state of alternating
filled and empty sites as in \Fig{fullspectrum},
as seen for example, in \Fig{edge000}(a). 
This way the QZD vacuum state acquires a non-trivial
entanglement strucure. Similarly, the baryon,
while attracted to the boundary, has significant spatial
extent. As such, from a symmetry perspective, it can assume any
flavor symmetry label that derives from the combination of three
particles. In terms of \SU{3} symmetry sectors this also permits
octets $(11)$ aside the singlet $(00)$ and $(30)$ [cf. \Eq{baryon:space}].
Hence baryons (and also antibaryons) also exist in the octet
representation $(11)$.  In order to get an octet meson then, the
simplest way to achieve this, is via an octet baryon with a
singlet antibaryon or vice versa. Given that the octet meson
splits $(b\bar{b})$ across the boundaries, the same
may therefore also be expected for the simpler
siuation of the meson singlet.

\section{Summary and Outlook}

In this work, we considered ``QZD", a \Zthree gauge theory with
three massive flavors of fermions in $1+1$ dimensions. We argued
that thanks to the periodicity of Gauss' law, it provides a
unique opportunity to study ``color" neutral isolated hadrons.
Using state of the art tensor network simulations that take
advantage of the full $\U{1}\times \SUf{3}$ global symmetries,
we determined the low-lying symmetry resolved spectrum of the
theory for different masses. We identified two special points,
where level crossing happens between the different symmetry
sectors and that most probably correspond to special theories.
We then confirmed that this system is in a confining phase by
verifying a striking feature of confinement in $1+1$ dimensions:
the small coupling expansion of hadrons is non analytic in
$g^2$, and starts at order $(g^2)^{2/3}$. We also studied the
spatial distribution of the different excitations in our system.
In particular, we confirmed that baryons are smaller at strong
coupling. We also directly observed how the lightest meson
transition from a single mesonic excitation to a pair of
baryon-anti-baryon.

This work lays the ground for many potential exciting studies in
$1+1$ dimensions and beyond.  A very interesting feature of this
model is that, thanks to the periodicity of Gauss law, the model
can be studied with a non-zero \textit{baryon} chemical
potential, in the ``color" neutral sector. Studying
thermodynamical quantities as a function of $\mu$ appears as an
interesting outlook. The system can also be put at finite
temperature. Studying properties of ``color neutral" baryons can
be envisaged. Extending on our analysis of how excitations are distributed in space open the door to performing $1+1$ dimensional ``tomography" 
of hadronic states. It could in particular inform on the size
dependence of baryons as a function of coupling strength. In
this direction, it appears that studying the model with two
light flavors and one heavier one, reducing \SUf{3} to
$\SUf{2}\times \U{1}$ is of merit. Real-time dynamics and scattering processes can also be studied, in the ground state as well as at non-zero density. Extending the model to $2+1$ dimensions and studying its phase diagram is also an interesting avenue. Finally, this model presents itself as a natural contender for analog as well as digital quantum computations. In $1+1$ dimensions, it is of the same complexity as the Schwinger model but gives access to different physics. In higher-dimensions, the gauge fields do not need to be truncated and reduce the complexity burden associated to bosonic degrees of freedom. 

\acknowledgements
The authors would like to thank J.~Barata for interesting discussions, and  S.~Mukherjee for discussions and for suggesting the name QZD.
A.F. and R.D.P.
were supported by the U.S. Department of Energy under contract 
DE-SC0012704 and by the U.S. 
Department of Energy, Office of Science, National Quantum 
Information Science Research Centers, Co-design Center
for Quantum Advantage (C$^2$QA), under contract number
DE-SC0012704.
A.W. was supported by the U.S. Department of Energy, Office of Science, Basic Energy Sciences, Materials Sciences and Engineering Division.
S.V.~was supported by the U.S. Department of Energy, Nuclear Physics Quantum Horizons program through the Early Career Award DE-SC0021892.

\appendix

\section{{Weak coupling expansion}}
\label{App:weakcoupling}
We derive here the weak coupling expansion \eqref{eq:weakcouplingmain} presented in the main text. The ``Coulomb" potential is obtained by solving Gauss law for a test charge $E(x) = \frac{1}{2}\mathrm{Sign}(x)$ and integrating. To obtain the correct small coupling expansion, it is crucial to remember we are using staggered fermion, so that the correct non-relativistic potential is obtained by integrating up to $x/2$,
\begin{equation}
V(x) = \frac{g^2|x|}{4} \ .
\end{equation}
(Equivalently, one could rescale $g^2\to g^2/2$ in Eq.~\ref{eq:discH}.)
The spectrum of the non-relativistic Hamiltonian is found by solving the associated non-relativistic Schr\"odinger equation
\cite{Fonseca:2006au,Fateev:2009jf,Ziyatdinov:2010vg,Zubov:2015ura}. As suggested by dimensional analysis, after rescaling $x\to y/(m g^2)^{1/3}$,
\begin{equation}
\left(\frac{g^4}{m}\right)^{1/3} \left(- \frac{1}{2}  \frac{d^2}{d y^2} + \frac{|y|}{4} \right)\psi(y) =  E_n\psi(y) \, .
\end{equation}
The function $\psi(y)$ are solutions to the Airy equations. Imposing continuity relations, we get 
\begin{equation}
\psi(y) \sim \mathrm{Ai}
\left(
\left(\frac{4 m}{g^4}\right)^{1/3}
\left(-2 E_n  + \frac12 g^2 y\right)
\right) \, .
\end{equation}
Valid solutions are split into symmetric and antisymmetric sectors. The symmetric sector is characterized by $\psi'(0)=0$ and contains the lowest-lying meson. The first zero of $\mathrm{Ai}'(-z)$ is $z\approx 1.01879$ \cite{NIST:DLMF}, which gives Eq.~\eqref{eq:weakcouplingmain}. 

This analysis is identical to that in the 't Hooft model \cite{tHooft:1974pnl,Fonseca:2006au,Fateev:2009jf,Ziyatdinov:2010vg,Zubov:2015ura}, which is a \SU{N} gauge theory in $1+1$ dimensions as $N\rightarrow \infty$, keeping the number of
quark flavors, $N_f$ fixed.  In this limit corrections to the gluon
propagator from the quark loop are suppressed by $\sim N_f/N_c$, and the gluon propagator remains $= 1/k^2$ for any value of the coupling constant.  In contrast, for QED in $1+1$ dimensions, in general the photon propagator is modified by
fermion loops.  However, in weak coupling, where $g^2/m^2 \rightarrow 0$, corrections to the photon propagator from fermion loops 
are suppressed by $\sim (g^2/m^2)^{1/3}$, and so can be neglected.

\section{{Symmetry labels}}
\label{app:symlabels}

The Hamiltonian in \Eq{discH} preserves particle
number and is fully symmetric in its $N_f =3$ fermionic flavors.
Hence it has $\U{1}_{\rm N} \otimes \SU{3}_{\rm flavor}$
symmetry. We fully exploit these symmetries in our numerical
simulations by utilizing the QSpace tensor library
\cite{Wb12_SUN,Wb20,qspace4u}.
Accordingly, we can differentiate all eigenstates
according to these symmetry sectors.

We specify symmetry labels
in terms of the tuple of three integer values
\begin{eqnarray}
   q &\equiv& (q_0; q_1, q_2) \equiv (q_0; q_1 q_2)
\label{eq:symlabels}
\end{eqnarray}
where $q_0 \in \mathbb{Z}$ specifies the total number of particles
relative to half-filling, and $(q_1,q_2) \equiv (q_1 q_2)$
specifies the \SU{3} multiplet.
The latter are based on the standard multiplet labels
for \SU{N} that directly specify the respective
Young tableaux \cite{Young30,Cahn84}.
This requires two labels $q_1,q_2 \geq 0$
for an \SU{3} multiplet
which specify a Young tableaux of two rows,
\begin{eqnarray}
   \includegraphics[width=0.7\linewidth]{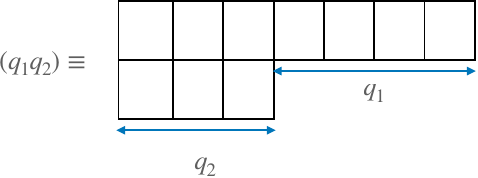}
\end{eqnarray}
where $q_1$ and $q_2$ indicate the offset of extra boxes
per row, starting from the top. This concept generalizes
to general \SU{N} \cite{Cahn84} with $N-1$ rows there.
E.g, for \SU{2}, $q_1=2S$.
Completely filled columns of $N$ boxes represent singlets
and can be skipped from the tableau.

\subparagraph{Local state space}

With the symmetries above, all $2^3=8$ states of a single
site are organized into symmetry multiplets as follows:
the completely filled state has symmetry labels $(3/2;00)$,
the completely empty state $(-3/2;00)$.
Half-integer particle numbers here is simply 
due the definition of subtracting half-filling $n_0=3/2$,
and has no further relevance otherwise
The same also holds for blocks containing
an odd number of sites.
The three states with only one particle transform in the defining 
representation of \SU{3}, hence represent the combined symmetry
multiplet $(-1/2;10)$. Conversely, removing a particle
from the completely filled state transforms in the
dual to the defining representation. 
Hence these states
represent the symmetry multiplet $(1/2;01)$.
In their union, $1+1+3+3=8$, this exhausts the local
state space.

We note that having half-integers for particle number is purely
   due to the definition `relative to half-filling'.
   In practice, via the tensor library QSpace \cite{Wb12_SUN}
   we use {\it twice} the particle number relative
   to half-filling, such that the symmetry label for the
   local particle number of a site relative to half-filling is also an integer, having $2 n'\in \{-3,-1,1,3\}$
   for a single site.

\subparagraph{Examples for \SU{3}} The defining representation
has symmetry labels $(10) \equiv {\bf 3}$, and its dual
$(01) \equiv \bar{\bf 3}$. The `spin' operator
transforms in the adjoint representation $ (11) \equiv {\bf 8}$
(octet), 
\begin{equation}
   {\bf 3} \otimes \bar{\bf 3} \equiv (10)\otimes (01) = (00) + (11)
\text{ ,}
\end{equation}
with $(00)\equiv {\bf 1}$ the scalar representation (singlet).
This also represents the symmetry labels of a single
particle-hole excitation (cf. meson).
Note that this is completely analogous to \SU{2}
where $\frac{1}{2} \otimes \frac{1}{2} = 0 + 1$,
with $S=1$ the \SU{2} spin operator.

Two particles transform in the combined space
\begin{equation}
   {\bf 3} \otimes {\bf 3} \equiv (10) \otimes (10) = (20) + (01)
\text{ ,}
\label{eq:two_particle}
\end{equation}
with $(20)\equiv  {\bf 6}$
the symmetric, and $(01) \equiv \bar{\bf 3}$ the antisymmetric subspace.
Three particles like the baryon transform in the combined space
\begin{eqnarray}
   {\bf 3} \otimes {\bf 3} \otimes {\bf 3}
   &\equiv& (10) \otimes (10) \otimes (10) \notag \\
   &=& (00) + (11)^2 + (30)
\text{ ,}\label{eq:baryon:space}
\end{eqnarray}
where superscript indicates multiplicities,
and $(30) \equiv {\bf 10}$ is a fully symmetric multiplet.
Dual representations are simply given by
$q=(q_1 q_2) \to \bar{q}=(q_2,q_1)$.
Hence all ireps with $q_1=q_2$ are self-dual,
while all others are not.

We emphasize that the specification of an
irreducible representation (irep) for \SU{N>2} via the single label
of its multiplet dimension only is generally insufficient
because it is not unique. For example for \SU{3}, the ireps
(40) and (21) accidentally share the same multiplet
dimension $d=15$ together with their respective duals
(04) and (12).

\section{{DMRG convergence}}

We use DMRG \cite{White92,Schollwoeck11}
in the fermionic setting where we fully exploit
the \SU{3} flavor symmetry for the
sake of numerical efficiency \cite{Wb12_SUN,Wb20,qspace4u}.
Data such as in \Fig{fullspectrum} was obtained
by simultaneously targeting several low energy multiplets
(cf. \App{symlabels}):
this included 4 multiplets in $(0,00)$, and one multiplet
in each of $(\pm 3,00)$, $(0,11)$, and $(3,11)$, 
i.e., a total of $8$ multiplets, or equivalently, 
$6 + 2\times 8 = 20$ states.

\begin{figure}[tbh]
\begin{center}
\includegraphics[width=0.8\linewidth]{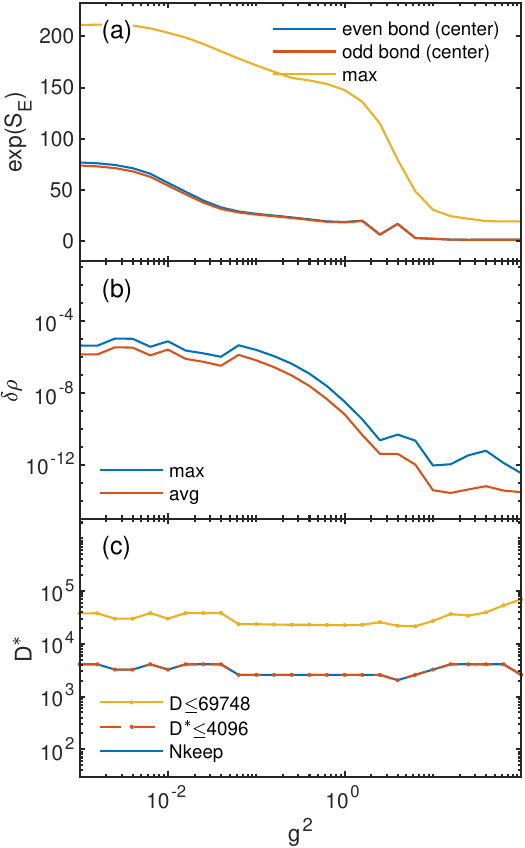}
\end{center}
\caption{
   DMRG convergence analysis for the data in \Fig{fullspectrum}
   vs. QZ3 coupling $g^2$ (having $m=0.125$, $L=120$).
   In the simulations a total of 8 multiplets was targeted
   simultaneously: 4 multiplets in $(0,00)$,
   and one multiplet in each of $(\pm 3,00)$, $(0,11)$, and $(3,11)$, 
   corresponding to a total of $6 + 2\times 8 = 20$ states.
   (a) Exponentiated block entanglement entropy in the 
   system center, and also the overall maximum along chain
   (the entanglement profile is strongly asymmetric
   around the system center, because multiple states are targeted).
   (b) Maximum and average discarded weight for last 2-site DMRG sweep.
   (c) DMRG bond dimension in last sweep, keeping up to 
   $N_{\rm keep} = D^\ast \leq 4,096$ multiplets
   (corresponding up to $D \leq 69,748$ states).
} 
\label{fig:DMRG} 
\end{figure} 

The bond dimension in terms of $D^\ast$ multiplets was
usually ramped up uniformly in an exponential way,
increasing it by a factor of $2^{1/3} \sim 1.26$ for each full sweep.
By keeping up to $D^\ast = 4,096$ multiplets, this effectively
corresponded to keeping up to $D \sim 70,000$ states
[\Fig{DMRG}(c)]. Thus by fully exploiting
\SU{3} flavor symmetry, the effective bond dimension was
effectively reduced by an average factor of $\sim 17$
by switching to a multiplet-based description.
Bearing in mind, that the numerical cost of DMRG scales like
$\mathcal{O}(D^3)$, this implies a gain in numerical
efficiency by at least three orders of magnitude.

For the data in \Fig{fullspectrum}, overall,
this gave rise to a discarded weight of $\delta\rho \lesssim 10^{-5}$
as shown in \Fig{DMRG}, with the entanglement
entropy [\Fig{DMRG}(a)] and thus also the discarded
weight largest for small $g$.

\section{{Mapping to spin Hamiltonian}}

In this appendix, we provide the spin-chain equivalent of \Eq{discH}. We obtain in using a standard Jordan-Wigner transformation and provide it only to assist the interested reader.

We introduce $3\cdot N$ spin operators $\sigma^{x,y,z}_I, \sigma^{\pm}_I=1/2(\sigma_I^x\pm\sigma_I^y)$, labeling them with an index $I=(n-1)\cdot 3 + f$ which uniquely maps onto indices $(n,f)$ for the position in the lattice, $n$ and the flavor, $f$.  The Jordan-Wigner transformation becomes $\chi_I = \sigma^-_I \prod_{J=1}^{I-1} \sigma^z_J$, and generates the spin Hamiltonian 
\begin{alignat}{2}
&H = \sum_{n=1}^{N} \sum_{f=1}^{3}   \lr{-\frac{i}{2a} \sigma^+_{n,f} \sigma^-_{n+1,f} S^{n+1,f}_{n,f}  + \mathrm{h.c.}} \notag\\
&+ \sum_{x=n}^{N} \sum_{f=1}^{3}   \lr{\frac{m}{2} (-1)^n \sigma^z_{n,f}}  \\
&+ \frac{g^2}{2} \sum_{n=1}^{N} \left( \frac{2\pi}{3}\sum_{l=1}^n \left ( \lr{\frac{\sigma^z_{n,f}}{2} +\frac{(-1)^n}{2}}\,\textrm{mod}\,3\right)\right)^2 \notag ,
 \end{alignat}
where $S^{n+1,f}_{n,f} = \prod_{J=3n-3+f}^{3n+f}\sigma^z_J$ is a string of $\sigma^z_{n,f}$ operators arising from the multi-flavor Jordan-Wigner transform. Similar strings arise in mapping a \SU{3} gauge theory in $1+1$ dimensions onto spin variables \cite{Farrell:2022wyt}.

\section{Nonzero chemical potential}

We present in this appendix exploratory results of QZD at non-zero baryon chemical potential. They are of value for this work as they provide a completely independent determination of the baryon mass and provide a convincing cross-check of our numerical analysis. 
For context,the behavior of QCD at low temperatures and chemical potential
is directly relevant to the collision of heavy ions at moderate
energies \cite{MUSES:2023hyz}
and to the behavior of neutron stars as observed by
multimessenger astronomy \cite{Dietrich:2020efo}.
At non-zero quark chemical potential
$\chem_{\rm qk}$
the quark determinant in the Euclidean action is complex, and so direct numerical simulations using importance sampling are not possible. When $\chem_{\rm qk}< T$,
thermodynamics quantities can be computed in several ways,
including: expanding in a Taylor series in $\chem_{\rm qk}$
\cite{Borsanyi:2020fev,Bollweg:2022rps,Bollweg:2022fqq,Mitra:2022vtf,Mitra:2023csk};
analytic continuation from imaginary chemical potential
\cite{Ishii:2018vvc,Begun:2021nbf,Bornyakov:2022blw,Brandt:2022jwo};
reweighting techniques
\cite{Borsanyi:2021hbk,Borsanyi:2022soo}; strong coupling
expansions
\cite{Gagliardi:2019cpa,Philipsen:2019qqm,Kim:2020atu,Klegrewe:2020cnp,Philipsen:2021qji,Philipsen:2021vgp,Kim:2023dnq};
complex Langevin equations
\cite{Kogut:2019qmi,Sexty:2019vqx,Ito:2020mys,Seiler:2023kes};
approximate solutions of the Schwinger-Dyson equations
\cite{Isserstedt:2019pgx,Gunkel:2021oya,Bernhardt:2021iql}; and
the functional renormalization group
\cite{Gao:2020qsj,Dupuis:2020fhh,Isserstedt:2020qll,Fu:2021oaw,Chen:2021iuo,Ayala:2021tkm,Fu:2022gou,Otto:2022jzl,Fu:2023lcm,Bernhardt:2023ezo}.

As a first step we consider QZD at $\mu \neq 0$,
finding the ground state of 
\begin{equation}
  H_\mu = H_0 - \mu \sum_x n_x
\ ,
\end{equation} 
as a function of $\mu$,
with $H_0$ the Hamiltonian in \Eq{H:continuum}.
For this simulation we had used
the package ITensor \cite{itensor,itensor-r0.3}
without imposing any symmetry constraint. In this DMRG simulation we kept up to $600$ states.

\begin{figure}
\includegraphics[]{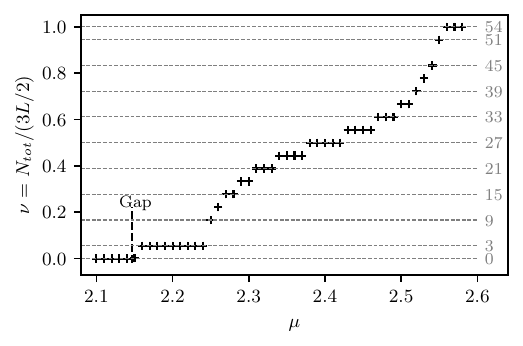}
\caption{
   The total number of particles as a function of the chemical
   potential, for $m=2$, $L=36$ and $g^2=1.466$.
   The vertical axis is scaled such that $\nu=1$
   is a completely filled system.
}
\label{fig:silverblaze}
\end{figure}

In Fig.~\ref{fig:silverblaze} we show the expectation value of the particle number as a function of $\mu$. 
It vanishes until $\mu = m_{(3,00)}$, where $m_{(3,00)}$ is the mass of the lightest baryon.  It is then constant until
it jumps again, to various multiples of three. 
That the number density vanishes until
$\mu > m_{(3,00)}$ illustrates "silver blaze" phenomenon \cite{Cohen:2004qp,Aarts:2015tyj}:
the ground state at $\mu=0$ remains the ground state of the grand canonical ensemble until the chemical potential exceeds
the mass of the lightest state which carries fermion number
It is an important consistency check
that $m_{(3,00)}$ determined from the silver blaze phenomenon
agrees with the direct calculation in Sec. \ref{sec:DMRG}.
That the number density only jumps to multiples of three follows
from gauge invariance under the local \Zthree symmetry:
baryons always carry $u$, $d$, and $s$ fermions in
common multiples.
This is in contrast to a \U{1} gauge theory, where as we
showed in the Sec. \ref{sec:introduction}, Gauss's law
excludes a nonzero value for the electric charge, or fermion number.  As $L \rightarrow \infty$, Fig.~\ref{fig:silverblaze}
would be a smooth curve, with $N/L$ a smoothly varying
function. 
For finite $L$, however, this is a series of steps
that increases in multiples of three, thus
guaranteeing a well-defined baryon number.
The absence of some multiples of three is an artifact
due to our resolution in $\mu$.
Note also that the fact the first plateau is larger than the other can probably be 
attributed to the open boundary as discussed
with \Fig{edge300} in the main text.

A more detailed study at finite chemical potential is left for future work.

\FloatBarrier

\bibliography{main.bib, manual.bib}

\clearpage

\end{document}